\documentclass[10pt,letterpaper,twocolumn]{article} 

\usepackage{ol2}
\usepackage[draft]{hyperref}
\usepackage{amsmath}

\begin{document}

\twocolumn[ 

\title{Reflectionless and invisible potentials in photonic lattices}


\author{Stefano Longhi}

\address{Dipartimento di Fisica, Politecnico di Milano and Istituto di Fotonica e Nanotecnologie del Consiglio Nazionale delle Ricerche, Piazza L. da Vinci 32, I-20133 Milano, Italy (stefano.longhi@polimi.it)}

\begin{abstract}
An arbitrarily-shaped optical potential on a discrete photonic lattice, which transversely drifts at a speed larger than the maximum one allowed by the light cone of the lattice band, becomes reflectionless. Such an intriguing result, which arises from the discrete translational symmetry of the lattice, is peculiar to discretized light and does not have any counterpart for light scattering in continuous optical media. A drifting non-Hermitian optical potential of the  Kramers-Kronig type is also an invisible potential, i.e. a discrete optical beam crosses the drifting potential without being distorted, delayed nor advanced.
\end{abstract}

 ] 

Reflection is ubiquitous in wave physics and is observed for a wide class of waves such as electromagnetic and particle waves \cite{r1}. In optics, reflection generally arises from a sharp change of the refractive index on a spatial scale of the order of the optical wavelength. However, it is known since long time that reflection can be avoided in certain dielectric media with specially-tailored refractive index profiles \cite{r2,r3}, even though the refractive index changes over a subwavelength spatial scale.  A class of such special graded-index profiles was introduced by Kay and Moses 
 in a pioneering work in the middle of the past century \cite{r2}. 
 Recently, wave reflection in engineered photonic structures has received a renewed interest, and new kinds of reflectionless potentials have been introduced, such as those based on parity-time ($\mathcal{PT}$) symmetry \cite{r4,r5,r6,r7,r8}, supersymmetry \cite{r9,r10,r11,r12}, and spatial Kramers-Kronig relations \cite{r13,r14,r15,r16,r17,r18}. In such structures, the refractive index is allowed to become complex, i.e. spatial regions with optical loss or gain are introduced. As compared to Kay and Moses potentials,  non-Hermitian potentials can be designed to be unidirectionally or bidirectionally invisible rather than simply reflectionless.\\ 
  \begin{figure}[htb]
\centerline{\includegraphics[width=8.4cm]{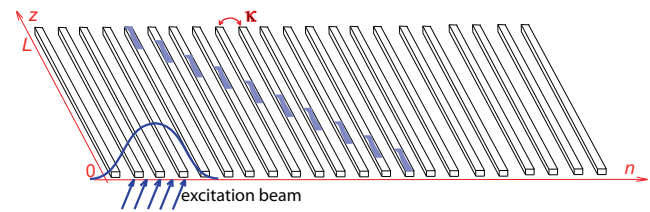}} \caption{ \small
(Color online) Schematic of light scattering on a waveguide lattice by a transversely drifting potential. The waveguide array is excited at the input plane $z=0$ by a tilted beam with positive group velocity. The scattering potential (dark areas) drifts on the lattice with a transverse velocity $v$ opposite to the group velocity of the injected beam. $n$ is the waveguide number, $\kappa$ the coupling constant between adjacent waveguides.}
\end{figure} 
 Integrated photonic structures provide a platform of major interest for molding the flow of classical and quantum light in unprecedented ways \cite{r19,r20,r21,r22}. Here light propagation is generally described by discrete equations, such as the discrete Schr\"odinger equation in lattice structures \cite{r19,r20,r21}, and it is generally referred to as discretized light \cite{r19}. Reflectionless potentials on a lattice, synthesized by a discrete version of supersymmetric quantum mechanics (also known as discrete Darboux transformation), have been introduced and experimentally demonstrated  in a few recent works \cite{r9,r23,r24,r25,r26}. In discrete optics, the continuous translational symmetry is broken and reflection is analogous to scattering by impurities in a periodic potential \cite{r21}. Like in a crystal, discrete translational symmetry drastically affects the linear  propagation of optical beams, with the appearance of allowed and forbidden energy bands and a wealth of related effects, such as anomalous refraction \cite{r27}, diffraction reversal and self-collimation \cite{r27,r28}, Anderson localization \cite{r29,r30}, negative Goos-H\"anchen shift \cite{r31}, etc.\\  
 In this Letter we disclose a rather intriguing property of wave scattering in discrete optics, namely the absence of reflection for discretized light by any arbitrarily-shaped {\it  moving} potential on a photonic lattice. The transparency effect is observed whenever the potential transversely drifts on the lattice faster than the largest speed allowed by the light cone of the lattice band. Such a result does not have any counterpart in continuous optical media, where a transverse shift of the potential at constant speed does not change its scattering properties. We also show that a drifting non-Hermitian potential of the Kramers-Kronig type \cite{r13} on a lattice is invisible, while it is reflective at rest.\\
  Let us consider propagation of discretized light waves in a waveguide lattice, which are scattered off by a transversely-moving optical potential (Fig.1). In the nearest-neighbor and tight-binding approximations, the evolution of mode amplitudes $c_n$ along the propagation distance $z$ in the lattice is described by the discrete Schr\"odinger equation  \cite{r19,r20,r21}
\begin{equation}
i \frac{dc_n}{dz}=- \kappa (c_{n+1}+c_{n-1})+V(n+vz) c_n
\end{equation}
    \begin{figure}[htb]
\centerline{\includegraphics[width=9cm]{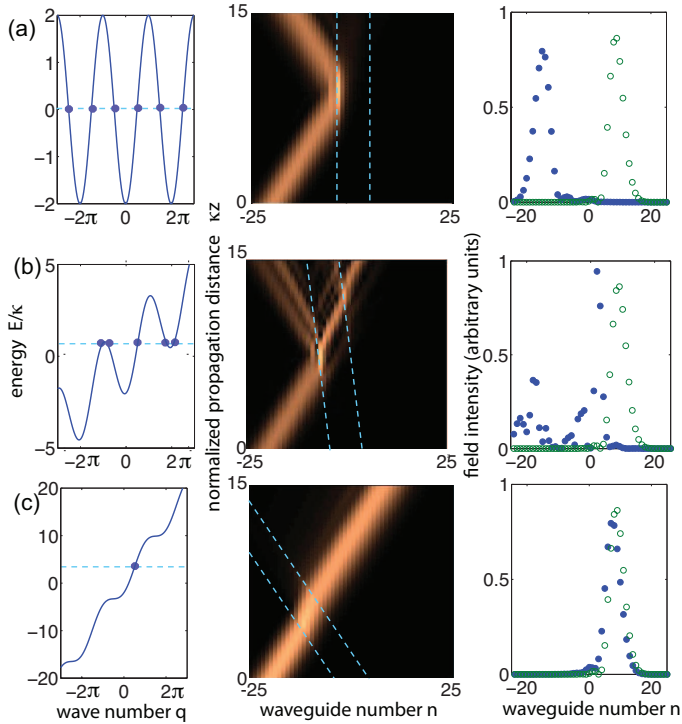}} \caption{ \small
(Color online) Left panels: Energy dispersion curves (solid lines) of a lattice in the moving reference frame $X=n+vz$, $Z=z$ for (a) $v=0$, (b) $v=0.4 \kappa<v_c$, and (c) $v=2.1 \kappa> v_c$, where $v_c=2 \kappa$ is the largest velocity of propagative waves allowed by the light cone of the lattice band. For a given energy $E$ determined by the incoming wave (dashed horizontal line), elastic scattering can couple waves with the same energy $E$ (circles). Distinct waves correspond to wave numbers $q$ which are not integer multiplies than $ 2 \pi$ each other. In (a) there are two distinct waves, in (b) five distinct waves, in (c) one wave solely. Central panels: numerically-computed evolution of a discretized Gaussian beam in the laboratory reference frame $(n,z)$ (snapshot of $|c_n(z)|$) for the three values of the drift velocities $v$ of left panels and for a rectangular scattering potential $V(n+vz)=V_0$ for $|n+vz|<d$, $V(n+vz)=0$ for $|n+vz|>d$ ($V_0= 3 \kappa$, $d=4$). The waveguide lattice comprises 50 waveguides and its length is $L=15 / \kappa$. The array is excited by a discretized Gaussian beam of spot size $w$ tilted at half the Bragg angle, i.e. $c_n(0) \propto \exp[-(n+n_0)^2/w^2+i (\pi/2) n]$ ($n_0=20$, $w=4$). The dashed lines in the plots schematically show the drifting rectangular potential. (c) Detailed beam intensity profiles at the output plane $z=L$ of the waveguide array (circles). The open circles show, for comparison, the beam intensity distributions that one would observe in the homogenous lattice, i.e. in the absence of the scattering potential.}
\end{figure} 
where $n=0 \pm 1, \pm 2$, .. is the waveguide number, $\kappa$ is the coupling constant between adjacent waveguides, $V=V(n)$ is the localized scattering potential, with $V(n) \rightarrow 0$ as $n \rightarrow \pm \infty$, and $v$ is the transverse drift velocity of the potential. In the following, we will assume $v>0$ and a forward-propagating optical wave incident from the left side of the scattering potential. The potential $V(n)$ can be either Hermitian, i.e. real and describing propagation constant offset of the waveguide mode, or non-Hermitian, i.e. complex and describing optical amplification or loss in addition to propagation constant offset of the mode. The continuous limit of Eq.(1), in which the discrete translation invariance of the lattice is lost,  is obtained from Eq.(1) by considering $n$ as a continuous variable and by letting $c_{n+1}+c_{n-1} \simeq 2 c(n,z)+ \partial^2_nc(n,z)$. Apart from an inessential constant potential, this yields the continuous Schr\"odinger equation for the amplitude $c(n,z)$
\begin{equation}
i \frac{ \partial c}{\partial z}=- \kappa \frac{\partial^2 c}{\partial n^2}+V(n+vz) c(n,z).
\end{equation}
Note that Eq.(2) also describes optical wave scattering from an inhomogeneous graded-index interface at grazing incidence \cite{r14,r32}. In such a continuous limit, wave scattering from the potential $V$ is independent of the transverse drift velocity $v$. This means that  a potential at rest which is not reflectionless can not be made reflectionless by just drifting it, and that in the continuous limit a reflectionless potential remains reflectionless when it drifts at an arbitrary speed. Such a general result readily follows from the Galileian invariance of the non-relativistic Schr\"odinger equation \cite{r33}, i.e. when considering the scattering problem in the reference frame 
\begin{equation}
 X=n+vz \; , \; Z=z
\end{equation}
where the potential is at rest. However, owing to breakdown of continuous translational symmetry the discrete Schr\"odinger equation (1) is not invariant under a Galileian transformation, and thus the invariance of the scattering process for a drifting potential is broken. In fact, in the new variables $X$ and $Z$, defined by Eq.(3), Eq.(1) takes the form
\begin{equation}
i \frac{\partial c}{\partial Z}=- \kappa \left[  c(X+1,Z)+c(X-1,Z) \right]-i v \frac{\partial c}{\partial X}+V(X)c.
\end{equation}
   \begin{figure}[htb]
\centerline{\includegraphics[width=8.9cm]{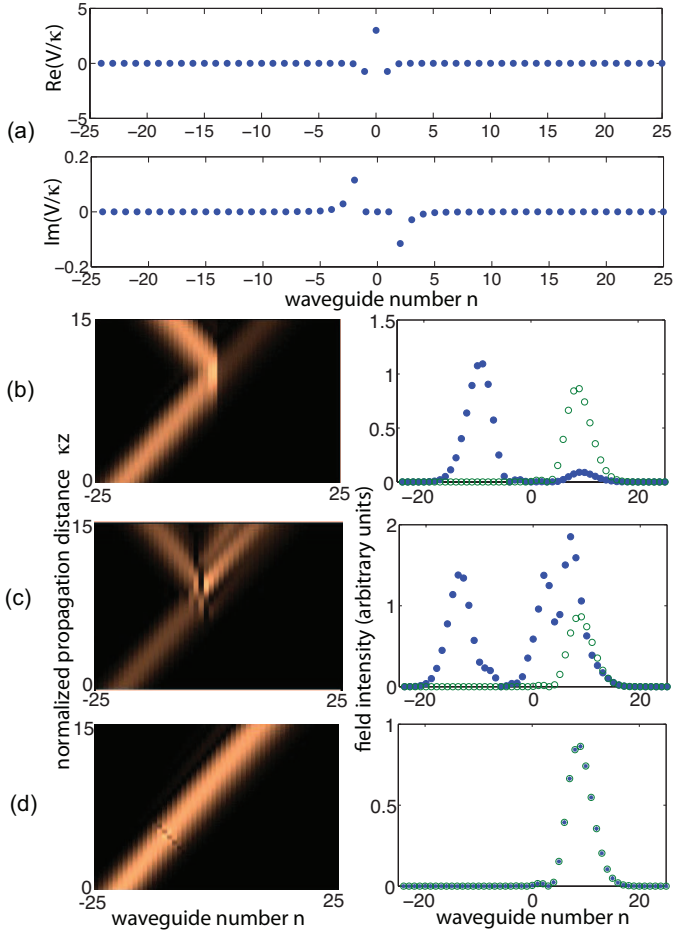}} \caption{ \small
(Color online) Scattering from a discrete Kramers-Kronig potential with drift velocity $v$. (a) Potential profile (behavior of the real and imaginary parts of $V(n)$). (b-d) Scattering of a tilted Gaussian beam for different drift velocities of the potential: (b) $v=0$, (c) $v=0.4 \kappa$, and (d) $v=2.1 \kappa$. The left panels in (b-d) show the numerically-computed evolution of the discretized beam in the laboratory reference frame $(n,z)$ (snapshot of $|c_n(z)|$), whereas the right panels show the detailed beam intensity profiles at the output plane $z=L$ of the waveguide array (circles). The open circles show, for comparison, the beam intensity distributions that one would observe in the absence of the scattering potential. Other parameter values are as in Fig.2. The sharp features in colormap arise from the discreteness of $n$.}
\end{figure} 
Unlike for the continuous Schr\"odinger equation \cite{r33}, the drift term $-i v (\partial c / \partial X)$ on the right hand side of Eq.(4) can not be removed by a gauge transformation, so that the scattering properties of the potential $V$ are modified by the drift term. In particular, for a velocity $v$ larger than the critical velocity $v_{c}=2 \kappa$ one can show that the potential $V$ becomes reflectionless. Note that $v_c=2 \kappa$ is the largest velocity for propagative discretized waves allowed by the light cone of the lattice band. To prove such a general property, let us  notice that in the reference frame $(X,Z)$ the scattering potential is at rest, so that scattering is elastic, i.e. it conserves the energy (i.e. propagation constant in the optical language). The value of the energy $E$ is determined by the incoming wave, which is scattered off by the potential. A solution to Eq.(4) can be then searched in the form $c(X,Z)=a(X) \exp(-i E Z)$, where $E$ is the energy eigenvalue and $a=a(X)$ satisfies the differential-difference equation
 \begin{equation}
 E a=-\kappa [a(X+1)+a(X-1)]-i v \frac{da}{dX}+V(X)a
\end{equation}
with $V(X) \rightarrow 0$ as $X \rightarrow \pm \infty$. Far from the scattering potential, i.e. for $X \rightarrow \pm \infty$, the scattering solutions to Eq.(5) are plane waves $a(X) \sim \exp(iqX)$ with wave number $q$ and energy
 \begin{equation}
 E(q)=-2 \kappa \cos(q)+vq.
 \end{equation} 
 The associated group velocity in the moving reference frame is given by $v_g= (dE/dq)=2 \kappa \sin q+v$, whereas in the laboratory reference frame $(n,z)$ it is given by $-2 \kappa \sin q$. Note that in the moving reference frame $(X,Z)$ the lattice dispersion curve, as given by Eq.(6), acquires a linear (ramped) term $vq$ in addition to the ordinary sinusoidal (periodic) term. For a drift velocity $v$ smaller than the critical velocity $v_c=2 \kappa$, one can find rather generally linearly independent plane waves with the same energy $E$ of the incoming wave and with wave numbers $q$ that are not integer multiplies than $ 2 \pi$ each other. Some of these distinct waves may correspond, in the laboratory reference frame $(n,z)$, to a group velocity opposite than the one of the incident wave [see, for example, the left panels of Figs.2(a) and (b)]. The scattering potential $V$ generally couples such waves, so that a reflected wave is typically observed in the laboratory reference frame. However, for a drift velocity $v$ larger than the critical velocity  $v_c$, regardless of the energy $E$ of the incoming wave there is only one wave number $q$ satisfying the relation (6); see left panel of Fig.2(c). Therefore reflection is here forbidden for energy conservation \cite{r34}, and the potential $V$ becomes reflectionelss. However, as compared to the case $V=0$, the scattering potential generally introduces an additional $q$-dependent phase shift to the transmitted plane wave, which results in beam distortion after potential crossing. In other words, an arbitrarily-shaped potential $V$ transversely drifting along the lattice at a speed $v>v_c$ becomes reflectionless but rather generally it is not invisible. As an example, Fig.2 shows the numerically-computed propagation of a discrete optical beam in an array made of 50 waveguides for a rectangular scattering potential [$V(X)=V_0$ for $|X|<d$, $V(X)=0$ for $|X|>d$] and for a few values of the transverse drift velocity $v$. In the laboratory reference frame $(n,z)$ the initial beam is Gaussian shaped and tilted at half the Bragg angle ($ q \simeq \pi/2$) from the entrance facet of the array, so as far from the scattering potential it propagates at the highest group velocity $v_c=2 \kappa$. For the potential at rest [$v=0$, Fig.2(a)] the wave packet is almost completely reflected, and the reflected wave packet has the same group velocity than the incident wave packet but reversed in sign. For a slowly-drifting potential [$v=0.4 \kappa$, Fig.2(b)], the wave packet is partially transmitted and partially reflected. Note that both reflected and transmitted beams break into some wave packets which propagate at different group velocities. Such a result can be readily explained by considering the energy diagram shown in left panel of Fig.2(b): for elastic scattering there are five distinct plane waves with the same energy $E=-2 \kappa \cos q+ q v=\pi v /2$ defined by the incoming wave packet, three with positive and the other two with negative group velocities (in the laboratory reference frame). The scattering potential transfers power from the incoming wave to the other four waves, resulting in break up of reflected and transmitted wave packets after the interaction.  For a fast-drifting potential [$v=2.1 \kappa$, Fig.2(c)], there is clearly no reflection and the wave packet fully crosses the scattering potential. However, as compared to the propagation in the homogenous lattice, i.e. with $V=0$, the transmitted wave packet is slightly distorted and delayed, as one can see from the right panel of Fig.2(c). Such a result indicates that, while the fast-drifting potential is reflectionless, it is not invisible. Note that in the continuous limit of the Schr\"odinger equation, i.e. for long wavelengths ($q \rightarrow 0$),  the dispersion relation (6) is replaced by $E(q) \simeq \kappa (q+v/ 2 \kappa)^2-2 \kappa-v^2 / 4 \kappa$, which is obtained by letting $\cos q \simeq 1-q^2/2$. Note that the dispersion curve is now parabolic, like in the stationary case $v=0$, the drift of the potential corresponding to a shift of the minimum of the parabola. Since the elastic scattering condition $E(q)=E$ shows two roots, corresponding to waves with opposite group velocities, reflection of the drifting potential is thus restored in the continuous (long wavelength) limit.\\
 Given the fact that any rapidly drifting potential (either Hermitian or non-Hermitian) on a lattice becomes reflectionless, a natural question arises, namely: are there potentials on a lattice that are also invisible? For wave scattering in continuous media, recent works have shown that a wide class of non-Hermitian potentials that satisfy the spatial Kramers-Kronig relations  can behave as unidirectionally or bidirectionally invisible potentials \cite{r13,r14,r15,r16,r17}. The reflectionless property of Kramers-Kronig optical potentials stems from the fact that a complex potential $V(X)$, for which the real and imaginary parts are related each other by a Hilbert transform, has a one-sided Fourier spectrum, and thus for one incidence side any scattered wave, at any order, has a wave number which can not become smaller than the one of the incident wave. In a continuous medium, this implies the absence of reflection. Regrettably, in a lattice with broken continuous translation symmetry and a periodic energy band the reflectionless property of Kramers-Kronig potentials at rest fails, because an increase of wave number can correspond to a sign change of the group velocity (Bragg scattering assisted by the crystal momentum). We now prove that, when the Kramers-Kronig potential drifts with a velocity $v>2 \kappa$, it becomes invisible. To this aim, let us consider a drifting potential at a speed $v> 2 \kappa$, so that there are not reflected waves. The asymptotic solution to Eq.(5) is thus of the form
 \begin{eqnarray}
 a(X) \sim \left\{  
 \begin{array}{ll}
 \exp(iqX) & X \rightarrow - \infty \\
 t(q) \exp(i q X) & X \rightarrow \infty
 \end{array}
  \right.
 \end{eqnarray}
 where $t=t(q)$ is the transmission coefficient. An invisible potential corresponds to $t(q)=1$. For an Hermitian potential $|t(q)|=1$ for power conservation, however the phase of $t$ can vary with $q$, leading to beam distortion, delay or advance [see for example Fig.2(c)]. Let us now assume that $V(X)$ is holomorphic in a half part of the complex plane $X=\xi+i \delta$ (either upper $\delta \geq 0$ or lower $\delta \leq 0$ half plane), i.e. $V(X)$ does not have poles nor branch cuts in a half part of the complex plane. This is equivalent to say that the real and imaginary parts of $V(X)$ are related each other by a Hilbert transform, i.e. by spatial Kramers-Kronig relations. For a meromorphic function, we assume that the sum of residues of $V(X)$ vanishes. Such a condition ensures that the potential vanishes at infinity sufficiently fast and the scattering states are plane waves \cite{r14}. The transmission coefficient can be determined by the method of imaginary displacement, i.e. by integration of Eq.(5) on the horizontal line $\Gamma$ of the half complex plane of analyticity of the potential. The parametric equation of the line $\Gamma$ is $X= \xi+i \delta_0$, with $\delta_0$ fixed and $ -\infty <  \xi < \infty$. The method is illustrated in Ref.\cite{r17}. Since the transmission coefficient $t$ does not depend on $\delta_0$, one can take the limit $|\delta_0| \rightarrow \infty$. For large displacement $|\delta_0|$ the potential $V(X=\xi+i \delta_0)$ on the line $\Gamma$ vanishes uniformly over $\xi$, so that there is not scattering at all and $t(q)=1$. This proves that the drifting Kramers-Kronig potential is invisible. We checked the above theoretical prediction by direct numerical simulations of coupled mode equations (2) in the laboratory reference frame. As an example, Fig.3 shows numerical results of beam propagation on a waveguide lattice scattered off by a Kramers-Kronig potential described by a meromorphic function with a fourth-order pole, namely $V(n)=V_0/(n+i \alpha)^4$ with $V_0=3 \kappa$ and $\alpha=1$. The real and imaginary parts of the potential are depicted in Fig.3(a). Note that the real and imaginary parts of the potential have opposite parity and the potential is $\mathcal{PT}$ symmetric, with balanced gain and loss regions around $n=0$. A discretized Gaussian beam, tilted at half the Bragg angle, excites the array at the entrance plane $z=0$ like in Fig.2. For a stationary potential, i.e. for $v=0$, the beam is almost completely reflected [Fig.3(b)], indicating that, contrary to what happens in continuous media \cite{r13}, in a lattice a stationary Kramers-Kronig potential is not reflectionless. A similar behavior is found for a drifting potential at a velocity smaller than the critical velocity [Fig.3(c)]. For a drift velocity $v>2 \kappa$, the potential does not reflect anymore the incident beam, which is transmitted as if there were not at all any potential [Fig.3(d)]: this is a clear signature that the potential is invisible.\par
 In conclusion, scattering of discretized light in photonic lattices by a drifting optical potential shows a very intriguing behavior which is related to the broken continuous translational invariance of the lattice. Unlike for scattering in a continuous medium, in which Galileian invariance ensures that a transverse drift does not modify the scattering properties of the potential, any arbitrarily shaped potential transversely drifting on a lattice at a speed larger than the maximum one allowed by the light cone of the lattice band becomes reflectionless. We also showed that non-Hermitian Kramers-Kronig potentials \cite{r13} drifting on the lattice behave as invisible potentials, while they are reflective at rest contrary to what happens in a continuous medium. Our results disclose a very different behavior of light scattering in continuous versus discrete optical media and provide a new twist for the synthesis of reflectionless and invisible potentials on an integrated photonic platform.

\newpage


 {\bf References with full titles}\\
 \\
 \noindent
1. J. Lekner, \textit{Theory of Reflection of Electromagnetic and Partilce Waves} (Kluwer, Dordrecht, 1987).\\
2. I. Kay and H. E. Moses, {\it Reflectionless transmission through dielectrics and scattering potentials}, J. Appl.  Phys. {\bf 27}, 1503 (1956).\\
3. L.V. Thekkekara, V.G. Achanta, and S. Dutta Gupta, {\it Optical reflectionless potentials for broadband, omnidirectional antireflection}, Opt. Express {\bf 22}, 17382 (2014).\\
4. Z. Lin, H. Ramezani, T. Eichelkraut, T. Kottos, H. Cao, and D.N. Christodoulides, {\it Unidirectional Invisibility Induced by $\mathcal{PT}$-Symmetric Periodic Structures}, Phys. Rev. Lett. {\bf 106}, 213901 (2011).\\
5. S. Longhi, {\it Invisibility in $\mathcal{PT}$-symmetric complex crystals}, J. Phys. A {\bf 44}, 485302 (2011).\\
6. L. Feng, Y.-L. Xu, W.S. Fegadolli, M.-H. Lu, J.E.B. Oliveira, V.R. Almeida, Y.-F. Chen, and A. Scherer, {\it Experimental demonstration of a unidirectional reflectionless parity-time metamaterial at optical frequencies},  Nature Mat. {\bf 12}, 108 (2013).\\
7. A. Mostafazadeh, {Perturbative unidirectional invisibility}, Phys. Rev. A {\bf 92},  023831 (2014).\\
8. M. Kulishov, H. F. Jones, and B. Kress, {\it Analysis of $\mathcal{PT}$-symmetric volume gratings beyond the paraxial approximation}, Opt. Express {\bf 23}, 9347 (2015).\\
9. S Longhi, {\it Invisibility in non-Hermitian tight-binding lattices}, Phys. Rev. A {\bf 82}, 032111 (2010).\\
10.  M.A. Miri, M. Heinrich, R. El-Ganainy and D.N. Christodoulides, {\it Supersymmetric optical structures}, Phys. Rev. Lett. {\bf 110}, 233902 (2013).\\
11. M.A. Miri, M. Heinrich, and D.N. Christodoulides, {\it SUSY-inspired one-dimensional transformation optics}, Optica {\bf 1}, 89 (2014).\\ 
12. S. Longhi, {\it Supersymmetric transparent optical intersections}, Opt. Lett. {\bf 40}, 463 (2015).\\
13. S.A.R. Horsley, M. Artoni, and  G.C. La Rocca, {\it Spatial Kramers-Kronig relations and the reflection of waves}, Nature Photon.{\bf 9}, 436 (2015).\\
14. S. Longhi, {\it Wave reflection in dielectric media obeying spatial Kramers-Kr\"{o}nig relations}, EPL {\bf 112}, 64001 (2015).\\
15. S.A.R. Horsley, C.G. King, and T.G. Philbin, {\it Wave propagation in complex coordinates}, J. Opt. {\bf 18}, 044016 (2016).\\
16.  S. Longhi, {\it Bidirectional invisibility in Kramers-Kronig optical media}, Opt. Lett. {\bf 41}, 3727 (2016).\\
17. S.A.R. Horsley and S. Longhi, {\it One-way invisibility in isotropic dielectric optical media}, Am. J. Phys. {\bf 85}, 439 (2017).\\
18. C.G. King, S.A.R. Horsley, and T.G. Philbin, {\it Perfect Transmission through Disordered Media}, Phys. Rev. Lett. {\bf 118}, 163201 (2017).\\
19. D. N. Christodoulides, F. Lederer, and Y. Silberberg, {\it Discretizing light behaviour in linear and nonlinear waveguide lattices}, Nature {\bf 424}, 817 (2003).\\
20. A. Szameit and S. Nolte, {\it Discrete optics in femtosecond-laser-written photonic structures}, J. Phys. B {\bf 43}, 163001 (2010).\\
21.  I.L. Garanovich, S. Longhi, A.A. Sukhorukov, and Y.S. Kivshar, {\it Light propagation and localization in modulated photonic lattices and waveguides}, Phys. Rep. {\bf 518}, 1 (2012).\\
22. T. Meany, M. Gr\"afe, R. Heilmann, A. Perez-Leija, S. Gross, M.J. Steel, M.J. Withford, and A. Szameit, {\it Laser written circuits for quantum photonics}, Laser \& Photon. Rev. {\bf 9}, 363 (2015).\\
23.V. Spiridonov and A. Zhedanov,  {\it Discrete reflectionless potentials, quantum algebras and q-orthogonal polynomials}, Ann. Phys. {\bf 237}, 126 (1995).\\
24. A. Szameit, F. Dreisow, M. Heinrich, S. Nolte, and A. A. Sukhorukov, {\it Realization of Reflectionless Potentials in Photonic Lattices}, Phys. Rev. Lett. {\bf 106}, 193903 (2011).\\
25. S. Odake and R. Sasaki, {\it Reflectionless potentials for difference Schr\"odinger equations}, J. Phys. A {\bf 48}, 115204 (2015).\\
26. M. Heinrich, M.-A. Miri, S. St\"utzer, S. Nolte, D.N. Christodoulides, and A. Szameit, {\it Observation of supersymmetric scattering in photonic lattices}, Opt. Lett. {\bf 39}, 6130 (2014).\\ 
27. T. Pertsch, T. Zentgraf, U. Peschel, A. Br\"auer, and F. Lederer, {\it Anomalous Refraction and Diffraction in Discrete Optical Systems}, Phys. Rev. Lett. {\bf 88}, 093901 (2002).\\
28. H. S. Eisenberg, Y. Silberberg, R. Morandotti, and J. S. Aitchison, {\it Diffraction Management}, Phys. Rev. Lett. {\bf 85}, 1863 (2000).\\
29.Y. Lahini, A. Avidan, F. Pozzi, M. Sorel, R. Morandotti, D. Christodoulides, and Y. Silberberg, {\it Anderson Localization and Nonlinearity in One-Dimensional Disordered Photonic Lattices}, Phys. Rev. Lett. {\bf 100}, 013906 (2008).\\
30. M. Segev, Y. Silberberg, and D.N. Christodoulides, {\it Anderson localization of light}, Nature Photon. {\bf 7}, 197 (2013).\\
31. M.C. Rechtsman, Y.V. Kartashov, F. Setzpfandt, H. Trompeter, L. Torner, T. Pertsch, U. Peschel, and A. Szameit, {\it Negative Goos-H\"anchen Shift in Periodic Media}, Opt. Lett. {\bf 36}, 4446 (2011).\\
32. S. Longhi, {\it Resonant tunneling in frustrated total internal reflection}, Opt. Lett. {\bf 30}, 2781(2005).\\
33. G. Rosen, {\it Galilean Invariance and the General Covariance of Nonrelativistic Laws}, Am. J. Phys. {\bf 40}, 683 (1972).\\
34. S. Longhi, {\it Robust unidirectional transport in a one-dimensional metacrystal with long-range hopping}, EPL {\bf 116}, 30005 (2016).\\



\begin{thebibliography}{99}




\bibitem{r1}
J. Lekner, {\it Theory of Reflection of Electromagnetic and Partilce Waves} (Kluwer, 1987).
\bibitem{r2}
I. Kay and H. E. Moses, J. Appl.  Phys. {\bf 27}, 1503 (1956).
\bibitem{r3}
 L.V. Thekkekara, V.G. Achanta, and S. Dutta Gupta, Opt. Express {\bf 22}, 17382 (2014).
 \bibitem{r4}
 Z. Lin, H. Ramezani, T. Eichelkraut, T. Kottos, H. Cao, and D.N. Christodoulides, Phys. Rev. Lett. {\bf 106}, 213901 (2011).
 \bibitem{r5}
S. Longhi, J. Phys. A {\bf 44}, 485302 (2011).
\bibitem{r6}
L. Feng, Y.-L. Xu, W.S. Fegadolli, M.-H. Lu, J.E.B. Oliveira, V.R. Almeida, Y.-F. Chen, and A. Scherer,  Nature Mat. {\bf 12}, 108 (2013).
\bibitem{r7}
 A. Mostafazadeh, Phys. Rev. A {\bf 92},  023831 (2014).
 \bibitem{r8}
 M. Kulishov, H. F. Jones, and B. Kress, Opt. Express {\bf 23}, 9347 (2015).
 \bibitem{r9}
 S Longhi, Phys. Rev. A {\bf 82}, 032111 (2010).
 \bibitem{r10}
 M.A. Miri, M. Heinrich, R. El-Ganainy and D.N. Christodoulides, Phys. Rev. Lett. {\bf 110}, 233902 (2013).
 \bibitem{r11}
M.A. Miri, M. Heinrich, and D.N. Christodoulides, Optica {\bf 1}, 89 (2014).
\bibitem{r12}
S. Longhi, Opt. Lett. {\bf 40}, 463 (2015).
\bibitem{r13}
S.A.R. Horsley, M. Artoni, and  G.C. La Rocca, Nature Photon.{\bf 9}, 436 (2015).
\bibitem{r14}
S. Longhi, EPL {\bf 112}, 64001 (2015).
\bibitem{r15}
S.A.R. Horsley, C.G. King, and T.G. Philbin, J. Opt. {\bf 18}, 044016 (2016).
\bibitem{r16}
S. Longhi, Opt. Lett. {\bf 41}, 3727 (2016).
\bibitem{r17}
S.A.R. Horsley and S. Longhi, Am. J. Phys. {\bf 85}, 439 (2017).
\bibitem{r18}
C.G. King, S.A.R. Horsley, and T.G. Philbin, Phys. Rev. Lett. {\bf 118}, 163201 (2017).
\bibitem{r19} 
D. N. Christodoulides, F. Lederer, and Y. Silberberg, Nature {\bf 424}, 817 (2003).
\bibitem{r20}
A. Szameit and S. Nolte, J. Phys. B {\bf 43}, 163001 (2010).
\bibitem{r21}
I.L. Garanovich, S. Longhi, A.A. Sukhorukov, and Y.S. Kivshar, Phys. Rep. {\bf 518}, 1 (2012).
\bibitem{r22}
T. Meany, M. Gr\"afe, R. Heilmann, A. Perez-Leija, S. Gross, M.J. Steel, M.J. Withford,
and A. Szameit, Laser \& Photon. Rev. {\bf 9}, 363 (2015).
\bibitem{r23}
V. Spiridonov and A. Zhedanov, Ann. Phys. {\bf 237}, 126 (1995).
\bibitem{r24}
A. Szameit, F. Dreisow, M. Heinrich, S. Nolte, and A. A. Sukhorukov, Phys. Rev. Lett. {\bf 106}, 193903 (2011).
\bibitem{r25}
S. Odake and R. Sasaki, J. Phys. A {\bf 48}, 115204 (2015).
\bibitem{r26}
M. Heinrich, M.-A. Miri, S. St\"utzer, S. Nolte, D.N. Christodoulides, and A. Szameit, Opt. Lett. {\bf 39}, 6130 (2014).
\bibitem{r27}
T. Pertsch, T. Zentgraf, U. Peschel, A. Br\"auer, and F. Lederer, Phys. Rev. Lett. {\bf 88}, 093901 (2002).
\bibitem{r28}
H. S. Eisenberg, Y. Silberberg, R. Morandotti, and J. S. Aitchison, Phys. Rev. Lett. {\bf 85}, 1863 (2000).
\bibitem{r29}
Y. Lahini, A. Avidan, F. Pozzi, M. Sorel, R. Morandotti, D. Christodoulides, and Y. Silberberg, Phys. Rev. Lett. {\bf 100}, 013906 (2008).
\bibitem{r30}
M. Segev, Y. Silberberg, and D.N. Christodoulides, Nature Photon. {\bf 7}, 197 (2013).
\bibitem{r31}
M.C. Rechtsman, Y.V. Kartashov, F. Setzpfandt, H. Trompeter, L. Torner, T. Pertsch, U. Peschel, and A. Szameit, Opt. Lett. {\bf 36}, 4446 (2011).
\bibitem{r32}
S. Longhi, Opt. Lett. {\bf 30}, 2781(2005).
\bibitem{r33}
G. Rosen, Am. J. Phys. {\bf 40}, 683 (1972). 
\bibitem{r34}
S. Longhi, EPL {\bf 116}, 30005 (2016).


 \end{thebibliography}
\end{document}